\documentclass[10pt,twocolumn,prl,aps,amssymb,amsmath,superscriptaddress]{revtex4}
\include{gyrokinetics-macros}
\usepackage{color,graphicx,ulem,bm}

\newcommand{\mbf}[1]{\mathbf{#1}}

\newcommand{\parl}{\parallel}
\newcommand{\pd}[2]{\frac{\partial #1}{\partial #2}}
\newcommand{\unit}[1]{\mathbf{\hat{#1}}}

\newcommand{\gyroR}[2][s]{\ensuremath{{\left< #2 \right>}_\mathbf{R_\mathrm{#1}}}}
\renewcommand{\eqref}[1]{Eq.\ (\ref{#1})}

\renewcommand{\gyroR}[1]{\ensuremath{{\left< #1 \right>}}}

\newcommand{\vpa}{v_{\parallel}}

\newcommand{\beq}{\begin{equation}}
\newcommand{\eeq}{\end{equation}}
\newcommand{\grad}{\nabla}

\newcommand{\smallness}{\rho_*}

\begin{document}

\title{Intrinsic rotation driven by non-Maxwellian equilibria in tokamak plasmas}

\author{M.\ Barnes}
\email{mabarnes@mit.edu}
\affiliation{Plasma Science and Fusion Center, Massachusetts Institute of Technology, Cambridge, MA 02138, USA}
\affiliation{Oak Ridge Institute for Science and Education, Oak Ridge, TN 37831, USA}
\author{F.\ I.\ Parra}
\affiliation{Plasma Science and Fusion Center, Massachusetts Institute of Technology, Cambridge, MA 02138, USA}
\author{J.\ P.\ Lee}
\affiliation{Plasma Science and Fusion Center, Massachusetts Institute of Technology, Cambridge, MA 02138, USA}
\author{E.\ A. Belli}
\affiliation{General Atomics, PO Box 85608, San Diego, CA 92168-5608, USA}
\author{M.\ F.\ F.\ Nave}
\affiliation{Associa\c{c}\~{a}o EURATOM/IST, Instituto de Plasmas e Fus\~{a}o Nuclear, Instituto Superior T\'{e}cnico, Technical University of Lisbon, Portugal}
\author{A.\ E.\ White}
\affiliation{Plasma Science and Fusion Center, Massachusetts Institute of Technology, Cambridge, MA 02138, USA}

\begin{abstract}

The effect of small deviations from a Maxwellian equilibrium on turbulent momentum transport
in tokamak plasmas is considered.  These non-Maxwellian features, arising from diamagnetic
effects, introduce a strong dependence of the radial flux of co-current toroidal angular momentum
on collisionality: As the plasma goes from nearly collisionless to weakly collisional, the flux reverses direction from radially inward to outward.  This indicates a collisionality-dependent transition from peaked to hollow rotation profiles, consistent with experimental observations of intrinsic rotation.

\end{abstract}



\maketitle

\paragraph{Introduction.}


Observational evidence from magnetic confinement fusion experiments indicates that
axisymmetric toroidal plasmas (tokamaks) that are initially stationary develop differential toroidal rotation even in the absence of external momentum sources~\cite{noterdaemeNF03,degrassiePoP04,bortolonPRL06,riceNF07,parraPRL12}.  This `intrinsic' rotation can depend sensitively on plasma density and current, with relatively small variations reversing the rotation direction from co- to counter-current~\cite{bortolonPRL06,duvalPPCF07,incecushmanPRL09,erikssonPPCF09,whitePoP13}.  
Conservation of angular momentum dictates that the intrinsic rotation 
is determined by momentum transport within the plasma.  
Since turbulence is the dominant transport mechanism in fusion plasmas~\cite{conwayPPCF08}, one must understand turbulent momentum transport to understand intrinsic rotation.

For the up-down symmetric magnetic
equilibria used in most experiments, the turbulent momentum transport for a non-rotating plasma 
can be shown to be identically zero~\cite{peetersPoP05,parraPoP11,sugamaPPCF11} unless one retains formally small effects that are usually neglected in analysis.
A self-consistent, first-principles theory has been formulated that includes these effects~\cite{parraNF11,parraPoP12}.  Of these effects, only radial variation of plasma profile 
gradients~\cite{diamondPoP08,wangPoP10,waltzPoP11,camenenNF11}
and slow variation of turbulence fluctuations along the mean magnetic field~\cite{sung13}
have been studied,
and these studies have not led to a theory
that explains the key dependences of intrinsic rotation in the core of tokamaks.  

In this Letter we consider the novel 
effect of small deviations from an equilibrium Maxwellian distribution of particle velocities on turbulent momentum transport.  These deviations arise naturally due to diamagnetic effects in plasmas with curved magnetic fields and pressure gradients~\cite{hintonRMP76}.  
They vary strongly with quantities
such as collisionality, plasma current, and the equilibrium density and temperature gradients in the plasma.  We show using direct numerical simulations that these non-Maxwellian
features, though small, introduce significant new dependences to the turbulent momentum transport.
We discuss the physical origins of the dependences and possible implications for
tokamak experiments.

\paragraph{Momentum transport model.}

Tokamak plasma dynamics typically consist of low amplitude, small scale turbulent fluctuations 
on top of a slowly evolving macroscopic equilibrium.  It is thus natural to employ a mean field
theory in which the particle distribution function, $f$, is decomposed into equilibrium, $F$, and 
fluctuating, $\delta f$, components.  The fluctuations are low frequency, $\omega$, 
relative to the ion Larmor
frequency, $\Omega$, and anisotropic with respect to the equilibrium magnetic field, with
characteristic scales of the system size, $L$, along the field and the ion Larmor radius, $\rho$,
across the field.
Expanding $f=f_0+f_1+...$, employing the smallness parameter $\smallness\doteq \rho/L \sim \omega/\Omega \sim \delta f / F \sim f_{j+1}/f_j\ll 1$, and averaging over the fast Larmor motion and over the fluctuation space-time scales, one obtains a coupled set of multiscale gyrokinetic equations for the fluctuation and equilibrium dynamics~\cite{sugamaPoP97,barnes,barnesPoP10b,parraPPCF10,abelRPP12}.

Typically only the lowest order system
of equations for $\delta f$ is considered.  However,
these equations have been shown to possess a symmetry that prohibits momentum transport in a non-rotating plasma~\cite{peetersPoP05,parraPoP11,sugamaPPCF11}.  Consequently, we include
in our analysis higher order effects arising from corrections to the lowest order (Maxwellian) equilibrium~\cite{parraNF11,parraPoP12}.  We limit our analysis to these non-Maxwellian corrections
because they are known to depend sensitively on plasma collisionality and current, which are
key parameters controlling intrinsic rotation in experiments~\cite{bortolonPRL06,duvalPPCF07,incecushmanPRL09,erikssonPPCF09,whitePoP13}.
We further simplify analysis by considering only electrostatic fluctuations and by performing the subsidiary expansions $\smallness\ll \nu_* \ll 1$ and $\smallness\ll  B_{\theta}/B \ll 1$, where $B$ is the magnitude of the equilibrium magnetic field, $B_{\theta}$ is the magnitude of the poloidal component, $\nu_*\doteq \nu_{ii} q R / v_{ti}$, $\nu_{ii}$ is the ion-ion collision frequency, $v_{ti}=\sqrt{2T_i/m_i}$ is the ion thermal speed, 
$T_i$ is the equilibrium ion temperature, $m_i$ is the ion mass, 
$q$ is a measure of the pitch of the magnetic field lines called the safety factor, 
and $R$ is the major radius of the torus.  These are good expansion parameters
in typical fusion plasmas.

Using $(\mbf{R},\varepsilon,\mu)$ variables, with
$\mathbf{R}$ the position of the center of a particle's Larmor motion, $\varepsilon=mv^2/2$ the particle's
kinetic energy, $\mu=mv_{\perp}^2/2B$ the particle's magnetic moment, $v$ the particle's speed,
and $\perp$ indicating the component perpendicular to the magnetic field,
the resulting equation for the fluctuation dynamics is
\beq
\begin{split}
&\frac{Dg_s}{Dt} + \left(\mbf{v}_{\parl}\cdot\nabla_{\parl} + \mbf{v}_{Ds}\cdot\grad_{\perp}\right)\left(g_s  - Z_s e \gyroR{\varphi}\pd{\hat{F}_s}{\varepsilon}\right) \\
&= -\gyroR{\mbf{\delta v}_{E}} \cdot \left(\grad_{\perp} g_s + \nabla \hat{F}_s 
+ \frac{m_s R \vpa}{T_s}F_{Ms}\nabla \omega_{\zeta,E}\right)\\
& + Z_s e \mbf{v}_{\parl}\cdot\nabla \hat{\Phi} \pd{g_s}{\varepsilon} + C_s,
\end{split}
\label{eqn:gk}
\eeq
where $g=\big<\delta f_1+\delta f_2\big>$, $\mbf{v}$ is particle velocity, the subscripts $\parl$ 
and $\perp$ denote
the components along and across the equilibrium magnetic field, $Ze$ is particle charge, 
$\varphi=\delta \phi_1 + \delta \phi_2$ and $\hat{\Phi}=\Phi_0+\Phi_1$ are fluctuating and equilibrium electrostatic potentials, $\hat{F}=F_{0} + F_{1}$, $D/Dt=\partial/\partial t + \mathbf{v}_E \cdot \nabla_{\perp}$,
$\gyroR{ . }$ is an average over Larmor radius at fixed $\mbf{R}$,  
$\mbf{v}_{Ds}$ is the drift velocity due to the Coriolis effect and due to curvature and inhomogeneity
in the equilibrium magnetic field, 
$\mbf{\delta v}_{E}=(c/B)\unit{b}\times\nabla_{\perp}\varphi$ and
$\mbf{v}_{E}=(c/B)\unit{b}\times\nabla\hat{\Phi}$ are $E\times B$ drift velocities, $\unit{\bm{b}}$
is the unit vector directed along the magnetic field,
$\omega_{\zeta,E}=-(c/RB_{\theta})(\partial\Phi_0/\partial r)$ is the toroidal rotation frequency due to $E\times B$ flow, $c$ is the speed of light,
the subscript $s$ denotes species,
and $C_s$ describes the effect of Coulomb collisions on species $s$.

Tokamak plasmas
are sufficiently collisional that the distribution of particle velocities is close to Maxwellian; 
i.e., $f_0=F_0=F_M$, with $F_M$ a Maxwellian.
Equilibrium deviations from $F_M$ 
are determined by the drift kinetic equation~\cite{hintonRMP76},
\beq
\mbf{v}_{\parl}\cdot \grad H_{1s} + \mbf{v}_{Ms} \cdot\grad F_{Ms} = C_{s}[H_{1s}],
\label{eqn:dk}
\eeq
where $H_1=F_1 + Ze\Phi_1 F_M / T$.
Finally, the electrostatic potentials are obtained and the system closed by enforcing
quasineutrality:
\begin{gather}
\label{eqn:qn1}
\sum_s Z_s \int d^3 \mbf{v} \left(g_s + \frac{Z_s e}{T_s}\left(\gyroR{\varphi}-\varphi\right)F_{Ms}\right) = 0,\\
\sum_s Z_s \int d^3 \mbf{v} \left(H_{1s} - \frac{Z_s e}{T_s}\Phi_1 F_{Ms}\right) = 0.
\label{eqn:qn2}
\end{gather}

With $g_s$ and $\varphi$ determined by Eqs.~(\ref{eqn:gk})-(\ref{eqn:qn2}),
the turbulent radial fluxes of energy, $Q$, and toroidal angular momentum, $\Pi$, are given by
\begin{gather}
Q_s = \left< \varepsilon_s \delta f_s \mbf{\delta v}_E \cdot \nabla r \right>_{\Lambda},\\
\Pi = \sum_s\left< m_s R^2 \delta f_s \left(\mbf{v}\cdot \nabla \zeta\right)\mbf{\delta v}_E \cdot \nabla r\right>_{\Lambda},
\end{gather}
where $\delta f = g + Ze(\left<\varphi\right>-\varphi)F_M/T$, $\zeta$ is toroidal angle, and
$\left<a \right>_{\Lambda}=\int dt\int d^3\mbf{r}\int d^3\mbf{v} \ a \ /\int dt \int d^3\mbf{r}$ is an integral over all velocity space and over a volume of width $w$ ($\rho\ll w\ll L$)
and time interval $\Delta t$ ($R_0/v_{ti} \ll \Delta t \ll \smallness^{-2} R_0/v_{ti}$)
encompassing several turbulence correlation lengths and times.

\paragraph{Results and analysis.}

We obtain the correction, $F_1$, to the equilibrium Maxwellian and the corresponding
electrostatic potential, $\Phi_1$, 
by solving Eqs.~(\ref{eqn:dk}) and~(\ref{eqn:qn2}) using
the drift kinetic code \texttt{NEO}~\cite{belliPPCF08}.  These quantities are then input to the
$\delta f$ gyrokinetic code \texttt{GS2}~\cite{dorlandPRL00}, which we have modified to 
solve Eqs.~(\ref{eqn:gk}) and~(\ref{eqn:qn1}) in the presence of $F_1$ and $\Phi_1$.
To calculate the `intrinsic' momentum flux that is present even for a non-rotating plasma,
we set the total toroidal angular momentum in a flux surface, which consists of diamagnetic 
and $E\times B$ contributions, to zero:
$\sum_s\left< (m_s R^2 \mbf{v}\cdot\nabla \zeta) f_s\right>_{\Lambda}
=\sum_s m_s n_s \left<R^2\right>_{\Lambda} \left(\omega_{\zeta,E} + \omega_{\zeta,d}\right)=0$,
with $\omega_{\zeta,d}=\sum_s \left<m_s R^2 (\mbf{v}\cdot\nabla\zeta) F_{1s}\right>_{\Lambda}/\sum_s m_s n_s\left<R^2\right>_{\Lambda}$ the diamagnetic contribution to the toroidal
rotation frequency and $n$ the number density.
The non-zero $E\times B$ rotation needed
to cancel the diamagnetic rotation breaks the symmetry
of the lowest order gyrokinetic equation and thus contributes to momentum transport, as do the 
non-Maxwellian equilibrium corrections we have included.

We consider a simple magnetic equilibrium with concentric circular flux surfaces known as
the Cyclone Base Case~\cite{dimitsPoP00}, which has been benchmarked extensively
in the fusion community.  The equilibrium is fully specified by the Miller model~\cite{millerPoP98},
with $q=1.4$, $\hat{s}\doteq \partial \ln q /\partial \ln r=0.8$, $\epsilon\doteq r/R_0=0.18$, 
$R_0/L_{n}=2.2$, and $R_0/L_{T}=6.9$, where: $r$ is the minor radius at the constant pressure
surface, or flux surface, of interest; $R_0$ is the major radius evaluated at $r=0$;
and $L_{n}$ and $L_{T}$ are the density and temperature gradient
scale lengths for both ions and electrons.  In order to obtain the gradient of $F_1$ appearing
in Eq.~(\ref{eqn:gk}),
we must additionally specify the radial dependence of these quantities.  For our base case,
we choose $R_0/L_n$, $R_0/L_T$, and $dq/dr$ to be constant in radius.

With these base case parameters specified, we conduct a series of simulations with kinetic electrons and deuterium ions,
varying $\nu_*$ and $\kappa\doteq R_0^2 d^2\ln T/dr^2$ about the baseline value of 
$\nu_* = 0.003$, $\kappa=0$.  Our \texttt{GS2} simulations use 32 grid points in the 
coordinate parallel
to the magnetic field (the poloidal angle), 12 grid points in $\varepsilon$, 37 grid points in
$\lambda=\mu/\varepsilon$, and 128 and 22 Fourier modes in the radial
and binormal coordinates, respectively.  The box size in both the radial and binormal coordinates
is approximately $125\rho_i$.

\renewcommand{\arraystretch}{2}
\begin{table}[t]
\caption{Collisionality dependence of $\omega_{\zeta,d}$}
\centering
\begin{tabular}{ | c || c | c | c | c | c | c | c |}
\hline
$\nu_*$ & 0.003 & 0.030 & 0.059 & 0.089 & 0.148 & 0.208 & 0.297 \\
\hline
$\dfrac{R_0\omega_{\zeta,d}}{v_{ti}}$ & 0.091 & 0.114 & 0.127 & 0.137 & 0.153 & 0.165 & 0.180 \\ [1ex]
\hline
$\dfrac{R_0^2}{v_{ti}}\dfrac{\partial\omega_{\zeta,d}}{\partial r}$ & -0.447 & -0.577 & -0.651 & -0.701 & -0.776 & -0.829 & -0.891 \\ [1ex]
\hline
\end{tabular}
\label{table:nuomega}
\end{table}

\begin{figure}
\includegraphics[height=2.2in]{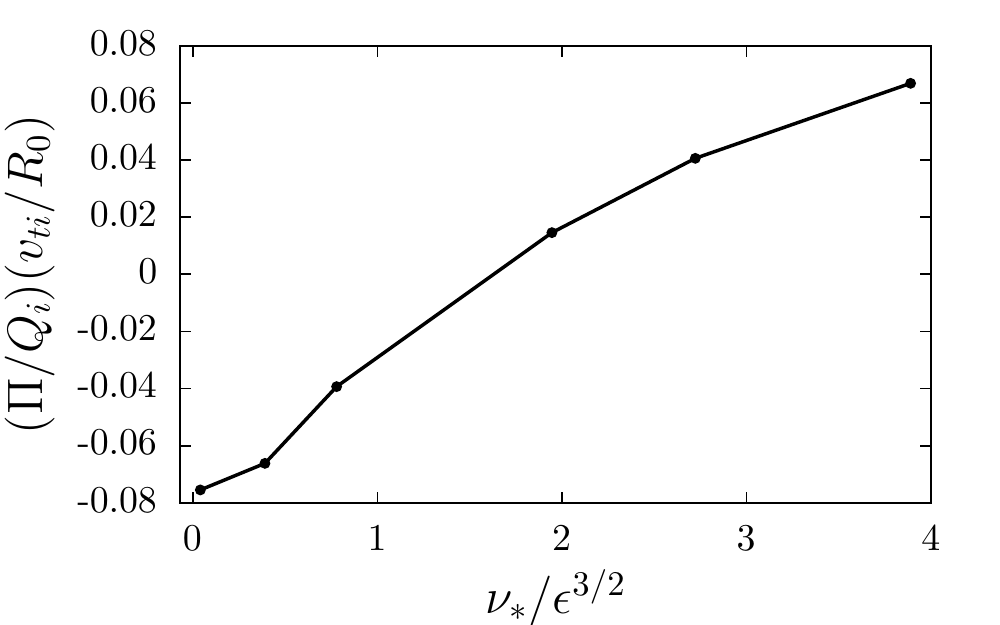}
\caption{Ratio of radial fluxes of ion toroidal angular momentum, $\Pi$, and energy, $Q_i$, vs. normalized ion-ion collision frequency, $\nu_{*}$.}
\label{fig:nupioq}
\end{figure}

The resulting $\Pi/Q_i$ values as a function of $\nu_*$ are shown in Fig.~\ref{fig:nupioq}.
We normalize $\Pi$ by $Q_i$, which is always positive, 
to remove any dependence of overall turbulence amplitude on collisionality.
Note that $F_1$, and thus $\omega_{\zeta,E}=-\omega_{\zeta,d}$, 
varies with collisionality, as indicated in Table~\ref{table:nuomega}.
For nearly collisionless plasmas, $\Pi/Q_i$ is negative, indicating a radially inward flux of
co-current angular momentum that would contribute to a centrally peaked rotation profile.
The ratio $\Pi/Q_i$ increases with $\nu_*$, passing through zero and becoming positive
when $\nu_* \sim \epsilon^{3/2}$.  For $\nu_* \gtrsim \epsilon^{3/2}$, the radially
outward flux of co-current angular momentum would contribute to a hollow rotation profile.

\begin{figure}
\includegraphics[height=2.2in]{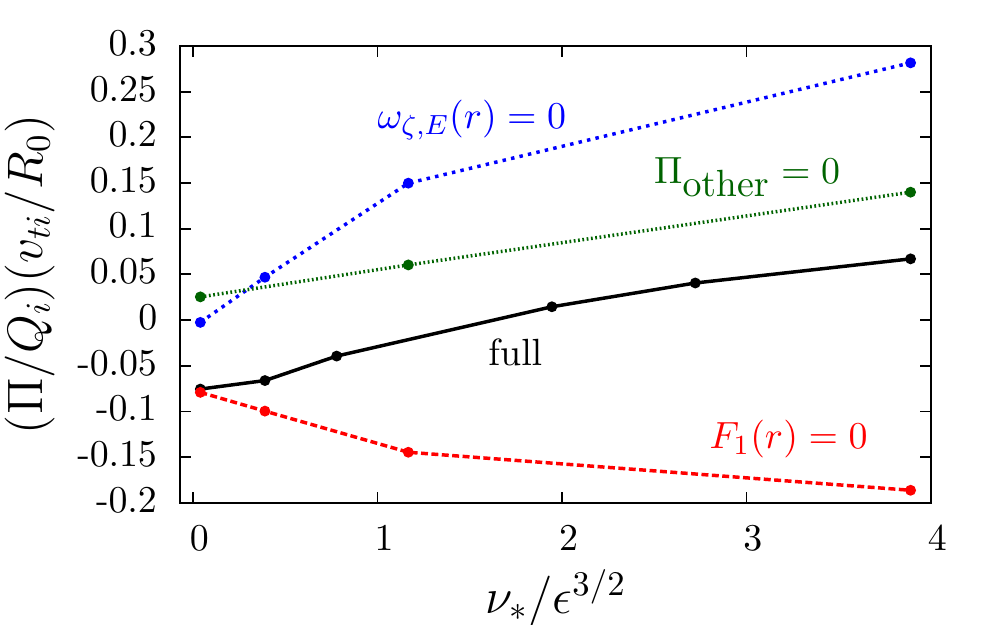}
\caption{Ratio of radial fluxes of ion toroidal angular momentum, $\Pi$, and energy, $Q_i$, vs. normalized ion-ion collision frequency, $\nu_{*}$ for: the base simulations (black), simulations with no $E\times B$ rotation to balance the diamagnetic rotation (blue), simulations with no correction, $F_1$, to the Maxwellian equilbrium (red), and simulations with $\Pi_{\textnormal{other}}=0$ (green).}
\label{fig:nupioqall}
\end{figure}

In Fig.~\ref{fig:nupioqall}, we show results from a series
of simulations in which we independently set the $E\times B$ rotation (including its derivative) and 
the diamagnetic effects, represented by $F_1$, to zero.  These are given by the 
blue and red curves, respectively.  We see that the $E\times B$ rotation causes
an inward momentum flux, with $\Pi/Q_i$ increasing in magnitude with $\nu_*$.  The
non-Maxwellian correction $F_1$ gives a $\Pi/Q_i$ that goes from slightly negative to large
and positive as $\nu_*$ is increased.  A partial cancellation between these effects gives the 
actual $\Pi/Q_i$.

To explore in more detail the origin of the sign reversal of $\Pi/Q_i$, 
it is convenient to express the ion energy flux 
in the diffusive form
$Q_i = -\chi_i dT_i/dr$,
and to decompose the momentum flux into 
diffusive, advective, and other pieces:
\beq
\begin{split}
\Pi &= -mR_0^2 \left(\pd{\omega_{\zeta,d}}{r}\chi_{\phi,d} + \pd{\omega_{\zeta,E}}{r}\chi_{\phi,E}\right) \\
&- mR_0 \left(\omega_{\zeta,d}P_{d} + \omega_{\zeta,E}P_{E}\right) + \Pi_{\textnormal{other}},
\label{eqn:pimodel}
\end{split}
\eeq
where $\chi_{\phi,d}$ and $P_{d}$ are diffusion and advection 
(commonly called `pinch') coefficients, respectively, for the diamagnetic rotation,
and $\chi_{\phi,E}$ and $P_E$ play the same roles for the $E\times B$ rotation.
The quantity $\Pi_{\textnormal{other}}$ accounts
for all other sources of $\Pi$ that arise due to $F_1(\omega_{\zeta,d}(r)=\omega_{\zeta,E}(r)=0)$;
e.g., the equilibrium parallel heat flow and other higher order velocity moments of $F_1$ 
will contribute to $\Pi_{\textnormal{other}}$.
Using the fact that $\omega_{\zeta,E}=-\omega_{\zeta,d}$ for a non-rotating plasma, we have
\beq
\Pi = -mR_0\left(R_0 \pd{\omega_{\zeta,d}}{r}\chi_{\phi,\textnormal{eff}} + \omega_{\zeta,d}P_{\textnormal{eff}}\right) + \Pi_{\textnormal{other}},
\label{eqn:pimodel}
\eeq
where $\chi_{\phi,\textnormal{eff}}=\chi_{\phi,d}-\chi_{\phi,E}$ and $P_{\textnormal{eff}}=P_d-P_E$.

Changing $\nu_*$ can alter $\Pi/Q_i$ in multiple ways.  First, the turbulent advection and diffusion
coefficients, $P_{\textnormal{eff}}/\chi_i$ and $\chi_{\phi,\textnormal{eff}}/\chi_i$, can be 
modified either directly by collisions or indirectly by the $\nu_*$-dependent rotation
and rotation gradient.  By independently varying $\nu_*$, $\omega_{\zeta,E}$, and $\partial\omega_{\zeta,E}/\partial r$ in \texttt{GS2} turbulence simulations with fixed $F_1$ and $\Phi_1$,
we found that such modifications of the advection and diffusion coefficients were minor.  
Furthermore, the turbulence type, characterized by the
dominant linear instability mechanism, remained the same (ion-temperature-gradient driven)
for all simulations.

With $\chi_{\phi,\textnormal{eff}}/\chi_i$ and $P_{\textnormal{eff}}/\chi_i$ approximately
independent of $\nu_*$, we see from Eq.~(\ref{eqn:pimodel}) that
the $\nu_*$ dependence of  $(\Pi-\Pi_{\textnormal{other}})/Q_i$ 
comes entirely from
the change of $\omega_{\zeta,d}$ and $\partial\omega_{\zeta,d}/\partial r$ with $\nu_*$, given
in Table~\ref{table:nuomega}.  In order to calculate $(\Pi-\Pi_{\textnormal{other}})/Q_i$,
we ran a series of simulations in which we used a modified $F_1$ that was constrained
to produce pure rotation so that $\Pi_{\textnormal{other}}=0$.  The results are shown
as the green curve in Fig.~\ref{fig:nupioqall}.  
We see that $(\Pi-\Pi_{\textnormal{other}})/Q_i$ is always positive
and increases approximately linearly with $\partial\omega_{\zeta,d}/\partial r$, 
as diffusion was found to
dominate over advection in these cases.
This indicates that equal and opposite diamagnetic and $E\times B$ rotations do not lead
to a complete cancellation of momentum transport~\cite{leePRL13}.
The increase in $(\Pi-\Pi_{\textnormal{other}})/Q_i$ with $\nu_*$
accounts for just over half of the total increase in $\Pi/Q_i$ over
the range of $\nu_*$ we have considered.  The rest of the increase, as well
as the negative offset needed to give the sign reversal in $\Pi/Q_i$ must come
from $\Pi_{\textnormal{other}}/Q_i$.

\begin{figure}
\includegraphics[height=2.2in]{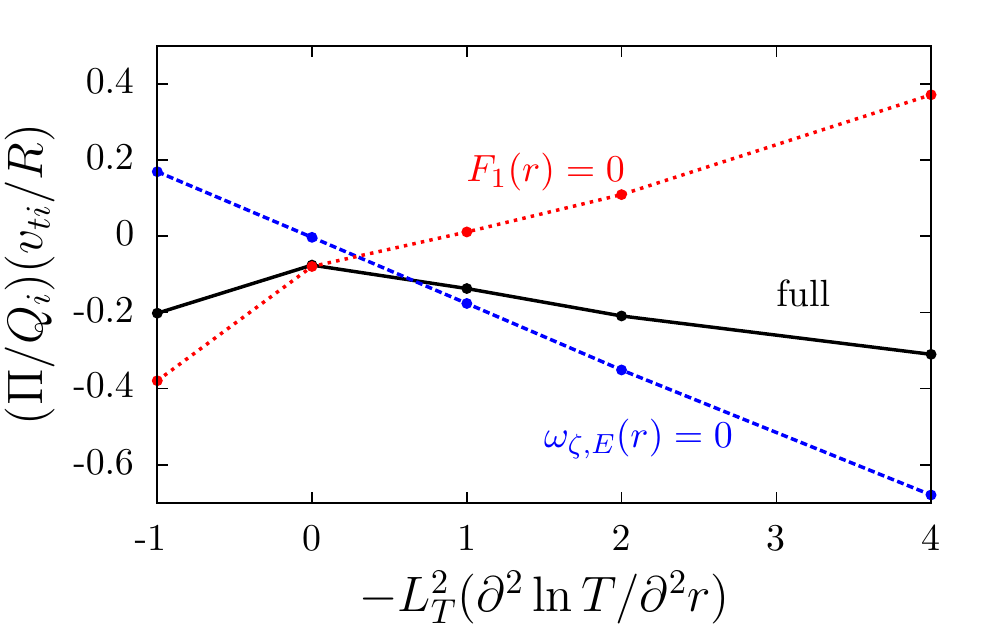}
\caption{Ratio of radial fluxes of ion toroidal angular momentum, $\Pi_i$, and energy, $Q_i$, vs. normalized second derivative of the logarithmic ion temperature for: the base simulations (black) and for simulations with no $E\times B$ flow to balance the diamagnetic flow (red) and no correction, $F_1$, to the Maxwellian equilibrium (blue).}
\label{fig:d2Tdr2pioq}
\end{figure}

To see how these results may be modified for different plasma profiles, we also conducted a 
series of simulations in which we fixed $\nu_*=0.003$ and varied $\kappa=R_0^2 \partial^2 \ln T/\partial r^2$.  Since the calculation of $F_1$ in \texttt{NEO} depends on $R_0/L_T$, varying $\kappa$ affects $\partial F_1/\partial r$ but not $F_1$ itself.  Consequently, $\partial \omega_{\zeta,d}/\partial r$ varies with $\kappa$ (see Table~\ref{table:kapomega}) while $\omega_{\zeta,d}$ itself remains fixed.
The change in $\Pi/Q_i$ with $\kappa$ is shown in Fig.~\ref{fig:d2Tdr2pioq}.  As was the case
in the $\nu_*$ study, the $E\times B$ and $F_1$ contributions to $\Pi/Q_i$ partially cancel,
though in this case each contribution independently changes sign with $\kappa$.  The net result
is a relatively weak variation of $\Pi/Q_i$ with no sign reversal.

\renewcommand{\arraystretch}{2}
\begin{table}[t]
\caption{Temperature profile dependence of $\omega_{\zeta,d}$}
\centering
\begin{tabular}{ | c || c | c | c | c | c |}
\hline
\ $-L_T^2\dfrac{\partial^2\ln T}{\partial r^2}$ \ & -1 & 0 & 1 & 2 & 4 \\ [1ex]
\hline
\ $\dfrac{R_0^2}{v_{ti}}\dfrac{\partial\omega_{\zeta,d}}{\partial r}$ \ & \ -1.116 \ & \  -0.447 \  & \  -0.105 \  & \  0.223 \  & \  0.835 \  \\ [1ex]
\hline
\end{tabular}
\label{table:kapomega}
\end{table}

\paragraph{Discussion.}

The sign reversal of $\Pi/Q_i$ shown in Fig.~\ref{fig:nupioq} suggests a transition from peaked
to hollow rotation profiles when $\nu_* \sim \epsilon^{3/2}$.  This is consistent with experimental
results, which show such transitions at similar $\nu_*$ values when density (proportional to $\nu_*$)  is increased or current (inversely proportional to $\nu_*$) is decreased~\cite{bortolonPRL06,duvalPPCF07,riceNF11,whitePoP13}.  Furthermore, our observation that the normalized turbulence diffusion and advection coefficients vary only minimally during the transition agree with recent experimental observations showing that the fundamental turbulence characteristics are unaltered
as the rotation reverses direction~\cite{whitePoP13}.

From Fig.~\ref{fig:nupioqall} and the analysis following Eq.~(\ref{eqn:pimodel}), it is evident that 
a combination of effects leads to the sign reversal of $\Pi/Q_i$.  However, the sign reversal
fundamentally originates from the $\nu_*$ dependence of $F_1$, which has been extensively studied 
and is the main concern of `neoclassical' theory (see, e.g., \cite{hintonRMP76,helander}).  
For $\nu_* \ll \epsilon^{3/2}$, known as the `banana' regime, all particle orbits are collisionless.
However,  for $\epsilon^{3/2} \ll \nu_* \ll 1$, known as the `plateau' regime, low energy particles 
that are trapped in the equilibrium magnetic well become collisional.  For a plasma
perfectly in the banana or plateau regimes, one can show that $F_1$, and thus our $\Pi/Q_i$,
becomes independent of $\nu_*$~\cite{hintonRMP76,helander}.  It is only when 
transitioning between these regimes that
$\Pi/Q_i$ varies with $\nu_*$.  So, while different profiles of quantities such as density, temperature,
and current may alter or eliminate the transitions with $\nu_*$ discussed above, they can only 
occur for $\nu_* \sim \epsilon^{3/2}$.

Finally, we reiterate that in our analysis we retained small terms (namely the diamagnetic effects that give rise to departures from a Maxwellian equilibrium distribution) in the multiscale gyrokinetic expansion, while we neglected other terms (radial profile variation, certain effects arising from the slow variation of fluctuations along the magnetic field, etc.) that may be of the same size.  There are two justifications for this.  First, 
if the fluctuation amplitudes and scales do not vary strongly with $B_{\theta}/B$, then the diamagnetic
effects considered here dominate so that our model is fully self-consistent~\cite{parraNF11,parraPoP12}.  Second, the
small effects we have neglected are not expected to have a particularly strong dependence on
collisionality.  Thus, while inclusion of these effects may provide an offset to the momentum transport,
we do not expect them to modify the variation of $\Pi/Q_i$ with $\nu_*$ presented here.

We thank J. Candy and P. J. Catto for useful discussions.
M.B. was supported by a US DoE FES Postdoctoral Fellowship, F.I.P. was supported by US DoE Grant
No DE-FG02-91ER-54109, and computing time was provided by the National Energy Scientific Computing Center, supported by the Office of Science of the U.S. Department of Energy under Contract No. DE-AC02-05CH11231.


\end{document}